\documentclass[journal]{IEEEtran}
\usepackage{cite}

\ifCLASSINFOpdf
   \usepackage[pdftex]{graphicx}
\else
\fi
\usepackage{algorithmic}
\usepackage{caption}
\usepackage{subcaption}

\usepackage{float}
\usepackage{stfloats}
\usepackage{url}

\usepackage{relsize}
\usepackage{tikz}
\usepackage{multirow}
\usepackage{multicol}
\def\textsubscript#1{\ensuremath{_{\mbox{\textscale{.6}{#1}}}}}
\usepackage{booktabs}
\usepackage{colortbl}
\usepackage{xcolor}
\usepackage{epstopdf}
\usepackage{mathtools}
\usepackage{multirow}
\usepackage{booktabs}
\usepackage{longtable}
\usepackage{ltablex}
\usepackage{makecell, longtable}

\usepackage{stfloats}

\newlength\mylen
\begin{document}
%
\title{Securing Optical Networks using Quantum-secured Blockchain: An Overview}
%
%
%

\author{Purva~Sharma,~\IEEEmembership{Student Member,~IEEE,}
        Vimal~Bhatia,~\IEEEmembership{Senior Member,~IEEE,}
        and~Shashi~Prakash, \IEEEmembership{Senior Member,~IEEE}
\thanks{P. Sharma and V. Bhatia are with the Signals and Software Group, Discipline
of Electrical Engineering, Indian Institute of Technology, Indore 453552 India (e-mail: phd1801202007@iiti.ac.in;  vbhatia@iiti.ac.in).}
\thanks{S. Prakash is with Photonics Laboratory, Institute of Engineering and Technology, Devi Ahilya University, Indore 452017, India (e-mail: spraksh@ietdavv.edu.in).}}
\maketitle

\begin{abstract}
Deployment of optical network infrastructure and network services is growing exponentially for beyond 5G networks. Since the uptake of e-commerce and e-services has seen unprecedented serge in recent months due to the global COVID-19 pandemic era, the security of such transactions in optical communication has gained much importance.  Optical fiber communication networks are vulnerable to several types of security threats, such as single point failure, wormhole attacks, and sybil attacks. Therefore, blockchain is a promising solution to protect confidential information against attacks and helps in achieving trusted network architecture by creating a distributed ledger platform. Recently, blockchain has received much attention because of its decentralized and distributed ledger technology. Hence, blockchain has also been employed to protect network against such attacks. However, blockchain technology's security relies on the platform of computational complexity, and because of the evolution of quantum computers, it will become insecure in the near future. Therefore, for enhancing blockchain security, research focus on combining quantum key distribution (QKD) with blockchain. This new technology is known as quantum-secured blockchain. The article describes the attacks in optical networks and provides a solution to protect network against security attacks by employing quantum-secured blockchain in optical networks. It provides a brief overview of blockchain technology with its security loopholes and focuses on QKD, which makes blockchain technology more robust against quantum-attacks. Next, the article provides a broad view of quantum-secured blockchain and presents the network architecture for future research and development of secure and trusted optical communication networks using quantum-secured blockchain.
\end{abstract}


%
\IEEEpeerreviewmaketitle

\section{Introduction}
%
%
%
%
\IEEEPARstart{o}{ptical} \small{n}etwork infrastructure and services are developing rapidly because of ever-increasing bandwidth-hungry applications (it will reach 4.8 zettabytes per year by 2022 \cite{cisco2018cisco}). The increase in security attacks such as single point failure, wormhole attacks, and sybil attacks make optical networks insecure and unreliable \cite{skorin2016physical}. Therefore, blockchain technology is used to avoid such types of attacks in optical networks by creating a decentralized environment \cite{kou2017blockchain}.

 Optical fiber communication networks are vulnerable to various types of security breaches, such as service disruption attacks and physical infrastructure attacks \cite[references therein]{skorin2016physical}.  
Service disruption attacks degrade performance by inserting interfering signals in the channel for jamming and alien-wavelength attacks. The physical infrastructure attacks, include single component failure, disaster attacks, and critical location attacks, that are physically damage the optical network infrastructure such as link or node component failure.  Nowadays, in control plane the software-defined network (SDN) controllers are installed, which provide logically centralized control to network operators and efficiently manages the network resources. However, an SDN controller is prone to single point failure, thereby making optical networks insecure. A malicious attacker may also use wormhole attacks and sybil attacks to disable networks by creating fake network resources \cite{kou2017blockchain}. The increasing security attacks in optical networks can cause huge data and revenue losses. Hence, problem related to optical network security is attracting more attention. Therefore, recently blockchain technology is being incorporated in optical network architecture to build trust between untrusted nodes in the network by monitoring the network resources in a distributed manner \cite{kou2017blockchain}. 

In software-defined optical networks, the blockchain technology has been introduced for providing trusted multi-controller routing, for implementing efficient failure recovery mechanism, and to ratify the quality of transmission (QoT) performance \cite[references therein]{fichera2020blockchain}. Recently, in \cite{fichera2020blockchain}, a novel framework based on blockchain technology has been proposed in order to provide trusted service level agreement (SLA) accounting in optical networks. Thus because of its attractive features and excellent performance, blockchain technology has become a hot topic of research.  
  
A blockchain is a distributed ledger or database based on cryptographic protection against malicious attacks. This technique allows users to share information among nodes in the network that do not trust each other \cite{ali2018applications}. Attractive features of blockchain, such as transparency, privacy, and accountability, make it reliable for variety of applications related to secure communication, smart contracts, healthcare, supply chain management, and other financial services. Blockchain technology was born with its most prominent application, namely, cryptocurrency Bitcoin \cite{nakamoto2008bitcoin}. Blockchain technology is more prevalent, and it is estimated that 10\% of global gross domestic product (GDP) will be saved on blockchain technology by 2025 \cite{marr2016blockchain}.

Blockchain security is based on one-way mathematical function and cryptographic algorithms, which is hard to hack. The conventional computers take too many years to break the blockchain security. However, as quantum computers are becoming a reality in future the blockchain security can be compromised \cite{kiktenko2018quantum}. Thus for improving the security of blockchain, post-quantum cryptography schemes are being designed \cite{fernandez2020towards}. However, such schemes are not strong enough to provide guaranteed security against quantum-attacks and are still in its infancy. Therefore, there has been a renewed interest in research on enhancing the security of blockchains by using quantum technologies.

The security of quantum communication relies on fundamental principles of quantum mechanics, i.e., Heisenberg uncertainty principle and quantum no-cloning theorem \cite{mailloux2015performance}. The uncertainty principle states that it is not possible to simultaneously measure the position and momentum of particles such as photons. Furthermore, any arbitrary unknown quantum states cannot be copied as stated by quantum no-cloning theorem. Quantum key distribution (QKD) is one of the most prominent applications of quantum communication. QKD generates secret keys between the end-users to encrypt and decrypt confidential information \cite{mailloux2015performance}, \cite{lo2014secure}. The secret key information is transmitted through quantum signal channel (QSCh); hence either sender or receiver can easily detect any security attack. Hence, QKD has potential to improve security of a blockchain network. Therefore, integrating QKD with blockchain is envisaged to pave way for new and secure technology termed as quantum-secured blockchain. Thus, this new and secure technology is a promising solution to improve security and performance of optical networks against malicious attacks.

In this article, we provide information related to types of attacks in optical communication networks. We provide a concise overview of blockchain technology and QKD with their basic defining process. We then discuss motivation behind integrating QKD with blockchain to form quantum-secured blockchain and explain its underlying process. After that, we focus on securing optical networks against threats by using quantum-secured blockchain. We then present a distributed quantum-secured blockchain optical network architecture and provide some challenges and opportunities for future research.


 

\section{Overview of Blockchain}
Blockchain is an innovative and unique technology for transferring and sharing confidential information among untrusted nodes in the network. It is a distributed database that consists of non-erasable records of information. In blockchain, the records are managed by a group of nodes in the network, and not by a single centralized authority, hence it is tamper-resistant \cite{belotti2019vademecum}. Blockchain security is based on two cryptographic tasks, i.e., a cryptographic hash function for encryption and digital signature for authentication, which makes blockchain more secure. In blockchain, each block is connected with its previous block using the previous block's hash value. Also, each node in the blockchain network has a copy of the ledger. Hence, if eavesdropper wants to break security of a blockchain, he/she has to solve a large mathematical problem of each node in the network at the same time, which is expensive and requires more computational power. Hence, the security of blockchain technology is almost unbreakable. In this section, we provide a short review of blockchain and its types \cite{ali2018applications}.

\subsection{Process of Blockchain}
\subsubsection{Blockchain Components}
A blockchain consists of following components for sharing and transferring confidential data between the end-users in the network.
\paragraph {Nodes} A node is a user or a computer who requests a transaction within the blockchain networks. Mainly there are two types of nodes in the blockchain networks, i.e., miner nodes and normal nodes. Miner nodes are those nodes which validate, authenticate, and verify the new blocks using consensus protocols in the network. Such nodes are the block generator. Normal nodes are those who have full information of blockchain content to maintain their database and cooperate with miners in the blockchain network.
\paragraph{Transaction} A transaction in blockchain network can be a financial data or any confidential information, depending on different types of applications.
\paragraph{Block} A block in a blockchain is like a record book. Each block consists of data (valid transactions), a hash value of block, a hash value of previous block, and a timestamp.
\paragraph{Merkle tree root hash}Merkle tree root hash value is the combination of repeating hash value of individual transactions, which are hashed again and again until a single hash value of a block is obtained.
\paragraph{Block hash} Block hash is a unique identity of block like a fingerprint. Once a block is created, its hash value is calculated by using hashing algorithms. It is beneficial when nodes in the network want to detect some changes in the block.
\paragraph{Previous block hash} Previous block hash value is always added to the hash of current block to create a chain and ensures immutability of the ledger.
\paragraph{Timestamp}A timestamp includes the creation time of the block, and it will monitor the creation time and update time of a block.
\paragraph{Genesis block} A genesis block is the initial block of blockchain. Each block in blockchain is sequentially added to the genesis block. This block is also known as \textit{block zero}.
\paragraph{Consensus protocol} The consensus protocol is a set of rules and regulations that helps in validating a new block. Different types of consensus protocols \cite{wang2019survey} have been designed for block validation, most widely used are proof-of-work (PoW), Byzantine Fault Tolerance (BFT), and proof-of-stake (PoS), discussed in \cite{belotti2019vademecum}.\\

\subsubsection{Working of Blockchain}
\begin{figure*}[ht!]
    \centering
    \includegraphics[scale=.55]{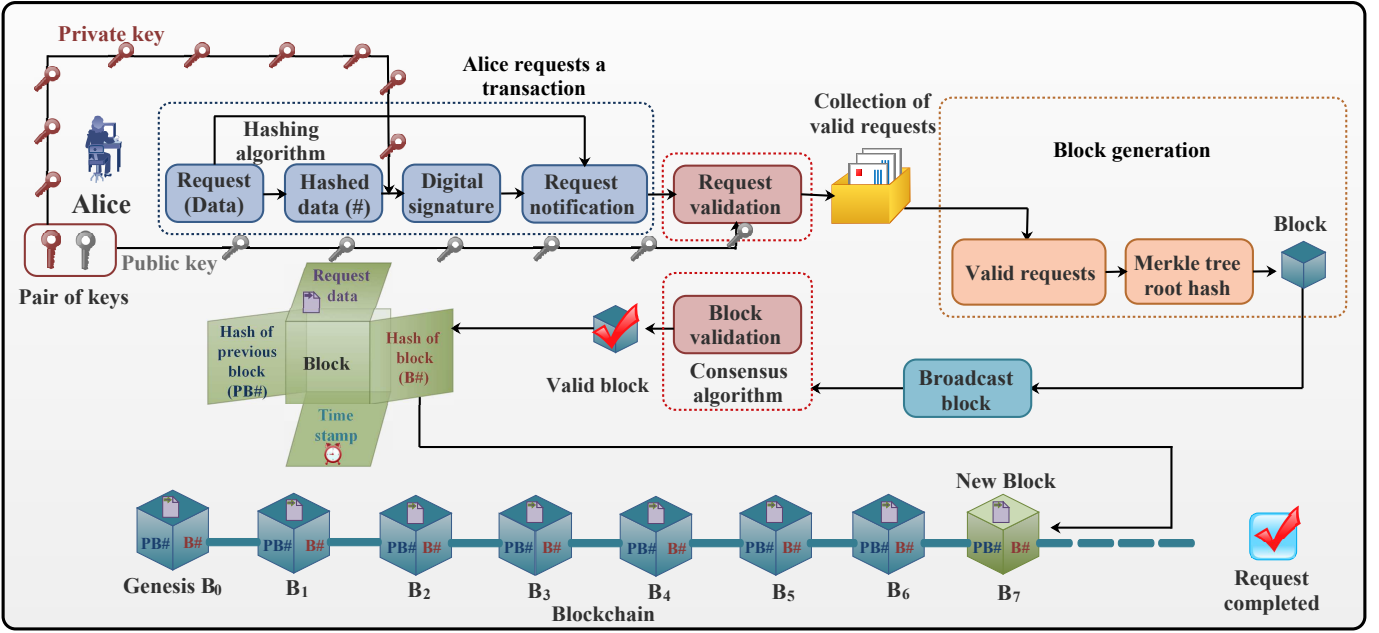}
   \caption{Process of Blockchain}
   \label{fig1}   
\end{figure*}

Fig. \ref{fig1} describes the blockchain process. Following steps explain the working of the blockchain technology \cite{ali2018applications}:
\paragraph{Transaction creation}Alice (sender) requests for a transaction. Before transmission, Alice uses cryptographic algorithms to encrypt and authenticate the transaction data. Alice first hashes the transaction data using hashing algorithms for data security. Each user (node) in the blockchain network generates a pair of keys, i.e., private key and public key using asymmetric cryptography. Alice uses her private key to sign the hashed data and generate a digital signature for authentication using elliptic curve cryptography.  A public key is used by the other nodes of blockchain network to authenticate the transaction data.

\paragraph{Broadcast and validation of transaction}After cryptography, both the transaction data and digital signature are broadcasted in a blockchain network. The nodes in a blockchain network validate the transaction by first decrypting the digital signature using the sender's public key for authentication and comparing the decrypted digital signature with hashed transaction data for integrity. Then the valid transactions are collected in a block.
\paragraph{Broadcast and validation of block} A block with valid transactions is broadcasted to selected miners in the network to generate a valid block. The miner uses consensus protocols to validate the block. After validation, the miner broadcasts a valid block in blockchain network and then it is added to the blockchain. At the end, the ledger of each node is updated in the blockchain network, as shown in Fig. \ref{fig1}. Hence, the request is completed.
\subsection{Features of Blockchain}
A blockchain has following characteristics \cite{ali2018applications} that make it attractive for various types of applications.
\subsubsection{Decentralization} In blockchain technology, transactions are not validated and authorized by a centralized authority as in a centralized system. The participation of centralized authority requires cost for maintenance and also creates performance issues. Therefore, the blockchain technology employs cryptographic algorithms to maintain records and authentication in a distributed environment.
\subsubsection{Immutability} All the valid transactions are stored in blocks of blockchain. Each block is connected with the previous block using its hash value generated by cryptographic hash function. If attacker tries to alter any previous record of block, it will affect all the succeeding blocks in the blockchain and easily detectable. Hence, the blockchain is immutable, i.e., the previously stored data has not been changed.
\subsubsection{Transparency}After validation process using consensus protocols, the valid transaction is added in block. Also, the ledger of each node is updated, and this process is publicly visible. Hence, it ensures the transparency and security of blockchain.
\subsubsection{Resistance to attack} All the nodes in the blockchain network hold identical copies of the records and update when the transaction is valid. Hence, the blockchain is resistant to attacks and information leakage.
 
\subsection{Types of Blockchain}
Blockchains are classified into three main categories \cite{ali2018applications}, namely, public blockchain, private blockchain, and consortium blockchain, according to different types of applications.
\subsubsection{Public Blockchains} A public blockchain is fully decentralized, where any participant can participate in creating new blocks and can access the content of blockchain. In public blockchain, anyone can keep a copy of blockchain and participate in validation process of new blocks. Such types of blockchain is also known as a permissionless blockchain because anyone can join without any permission. A public blockchain network consists of a large number of nodes, so it is resistant to malicious attacks. Additionally, each transaction has some processing fees, as an incentive for user who participates in the validating process. This makes the public blockchain more transparent and secure. Examples of public blockchains are cryptocurrency networks such as Bitcoin and Ethereum.
\subsubsection{Private Blockchains}
Private blockchain is also known as a permissioned blockchain, where every node is a member of a single organization or institute. In private blockchain, an authority can access the content of blockchain and permit other users to access the content. There is no processing fees of transaction in private blockchain, and it is similar to that of a centralized system; however, it is cryptographically secure.
\subsubsection{Consortium Blockchains} A consortium blockchain is a special type of private blockchain where a selected number of participants from multiple organizations can participate in the consensus process. A consortium blockchain helps in maintaining transparency between the involved organizations. Similar to a private blockchain, there are no transaction processing fees, hence has lower cost. A consortium blockchain is partially decentralized or tamper-proof. Example of a consortium blockchain is Hyperledger.


\section{Quantum Key Distribution}
In this section, we provide an overview of QKD technology for secure communication. QKD establishes secure connection between the end-users by generating and distributing secret keys over an insecure channel. For secret key generation and distribution, QKD requires QSCh and public interaction channel (PICh), as shown in Fig. \ref{fig2} \cite{mailloux2015performance}. QSCh is used to send quantum bits (qubits), i.e., encoded polarization photons between the end-users. PICh is used to transmit measuring-basis of qubits and verify the secret keys using post-processing methods. Additionally, a QKD protocol is used to generate a secret key between the end-users and ensure security against eavesdropping. Several QKD protocols are designed for secret key generation, discussed in \cite{scarani2009security}. The most widely used QKD protocol is Bennett and Brassard-84 (BB84) protocol proposed in 1984 \cite[reference therein]{scarani2009security}.
\subsection{QKD Process}
\begin{figure*}[ht]
    \centering
    \includegraphics[scale=.6]{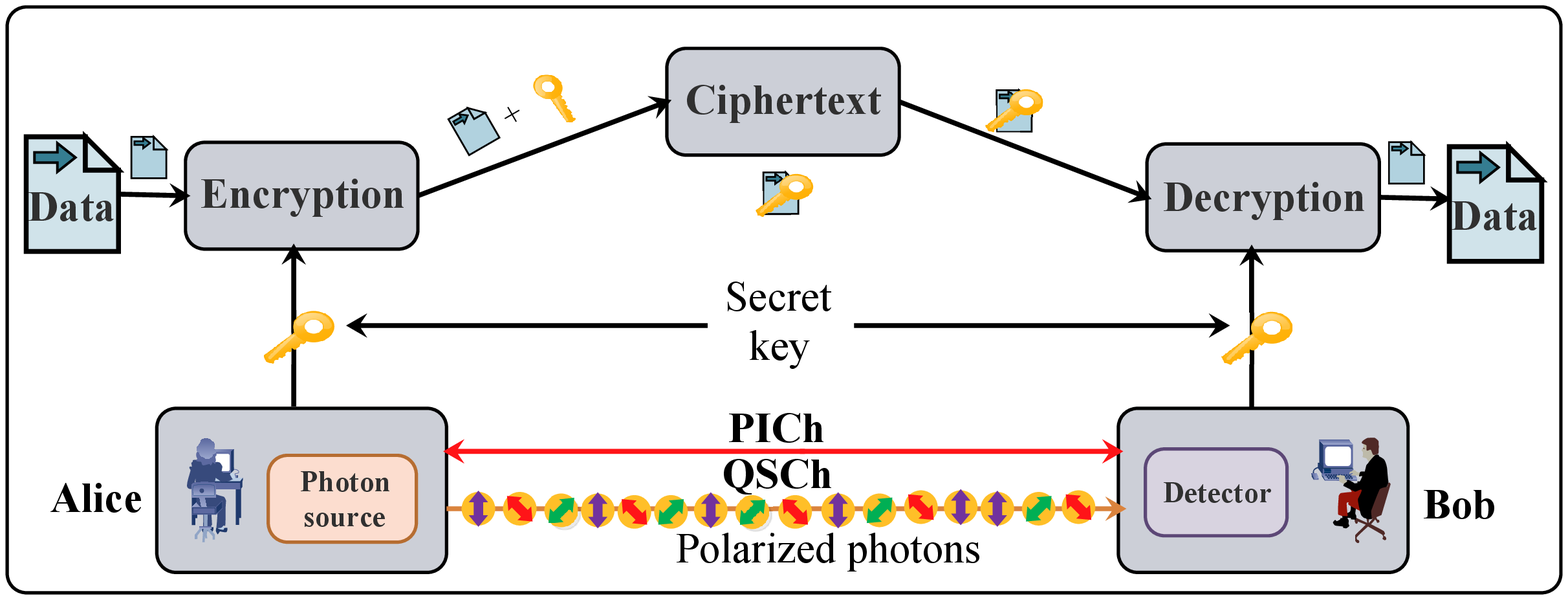}
   \caption{QKD Process \cite{mailloux2015performance}}
   \label{fig2}   
\end{figure*}

Fig. \ref{fig2} explains the QKD process for establishing secure communication between Alice (sender) and Bob (receiver). The following steps explain the process of a QKD system with BB84 QKD protocol \cite{mailloux2015performance}.

\begin{itemize}
\item Alice generates random string of bits and for each bit, she will randomly choose a basis, either rectilinear (two polarization states, i.e., $0^\circ$ or $90^\circ$) or diagonal (two polarization states, i.e., $+45^\circ$ or $-45^\circ$) with their polarization states. The random string of bits encoded with these polarization states are known as qubits. Alice then sends qubits to Bob through QSCh.
\item Bob receives the qubits from Alice and measures the received qubits with one of the randomly selected measuring bases, and obtains a string of all received qubits from measurement result. 
\item Alice and Bob exchange their measurement bases through PICh and compare with each other. After comparison, the qubits with different measuring bases are discarded. The remaining qubits corresponds to same measuring bases are decoded into a string of binary  bits known as \textit{Sifted key}.
\item A random substring of sifted key is exchanged and compared for parameter estimation and error correction between Alice and Bob via PICh.
\item Privacy amplification and authentication are performed, which reduces the information of remaining bits against the eavesdropping and generate a new shorter key known as \textit{Secret key}.
\item After secret key generation, encryption process starts. In this, the generated secret key encrypts the information transmitted by Alice and converts the information into ciphertext. Now, Bob uses the same secret key for decryption, i.e., convert the ciphertext into original information. In this way, Alice and Bob securely communicate with each other using QKD.
\end{itemize}

\section{Quantum-secured Blockchain}

Blockchain technology is strong enough to provide security within the blockchain network between the nodes by leveraging asymmetric cryptography and hashing algorithms. Asymmetric cryptography generates a pair of keys for providing security between the nodes and authenticate transactions by generating a digital signature. The most widely used digital signature schemes are Rivest, Shamir, Adleman (RSA), or elliptic curve cryptography \cite{fernandez2020towards}. Hashing algorithms also play a crucial role in providing security by hashing the transaction data and linking blocks of a blockchain by generating block hash values. However, security of both asymmetric cryptography and hash algorithms rely on computational complexity of certain mathematical functions that can be easily attacked by quantum computers in the near future \cite{kiktenko2018quantum}. Hence, blockchain releases all its security features and becomes insecure. If quantum attack-aware schemes are not designed to enhance the blockchain security then the existing and future blockchain networks will become vulnerable as it has been estimated that 10$\%$ of the global GDP will be stored on blockchain-based technology by 2025 \cite[reference therein]{kiktenko2018quantum}.

The post-quantum cryptography schemes \cite[reference therein]{fernandez2020towards} were proposed to overcome the problem of security in blockchains, however their security is questionable. Hence, not providing guaranteed security against threats. A most prominent way to provide complete security in blockchain against quantum-attacks is QKD. The security of QKD relies on the fundamental laws of quantum mechanics. QKD generates random secret keys between the authenticated users in the network using QSCh and PICh to encrypt confidential information. Hence, there is a huge research interest on protecting the blockchain network against quantum-attacks by integrating QKD in blockchains. Recently, a quantum-secured blockchain platform has been developed and experimentally demonstrated, which uses QKD for authentication and original BFT consensus protocol for validation \cite{kiktenko2018quantum}. The security of quantum-safe blockchain is practically realizable and scalable for different government and commercial services. However, a major drawback of the proposed quantum-secured blockchain is the use of consensus protocol. The limitation of BFT consensus protocol is that if a large number of non-operational nodes are present in the blockchain network, it becomes data-intensive. Hence, a  new quantum-secured consensus protocol is designed to limit the problem of traditional consensus protocol in \cite{sun2019towards}. However, not many protocols have been implemented to improve the security of blockchain networks. Therefore further research is urgently needed to design secure consensus protocols using quantum technologies.

\subsection{Process of Quantum Blockchain}
Quantum blockchain uses the same components as traditional blockchain, however a major difference being that instead of conventional cryptography and hashing algorithms, it utilizes quantum techniques to secure network against security breaches. Fig. \ref{fig3} shows the process of quantum blockchain. Following phases will explain the process of quantum blockchain in detail:
\begin{figure*}[ht]
    \centering
    \includegraphics[scale=.55]{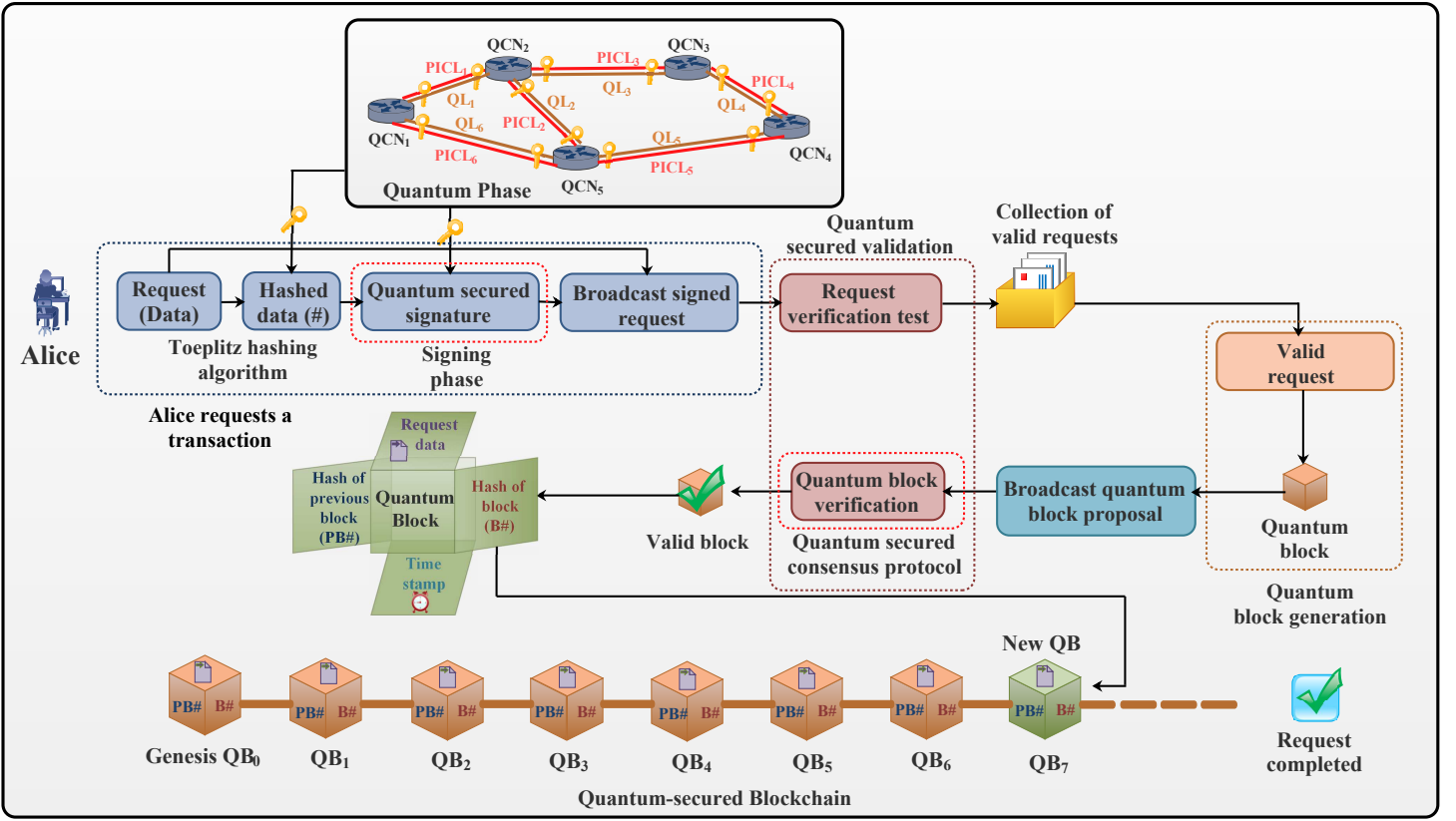}
   \caption{Process of Quantum-secured Blockchain}
   \label{fig3}   
\end{figure*}

\subsubsection{Quantum phase} In quantum phase, random secret keys between the two authenticated users in the network are generated by using QKD through QSCh and PICh. The generated secret keys are used for encryption and authentication. 
\subsubsection{Transaction proposal phase}Alice requests a transaction and hashed data by using hashing algorithms in the encryption phase. The most widely used scheme is Toeplitz hashing in which a Toeplitz matrix is generated by shared random keys between sender and receiver. This scheme along with one-time pad encryption, help in transferring transaction data securely. The generated secret keys using QKD are used in generating quantum-secured signature to sign a transaction in a signing phase. After the signing phase, the transaction data and the signature are broadcasted to the nodes in the quantum blockchain network.
\subsubsection{Transaction validation phase} In this phase, upon receiving a transaction data and signature, the blockchain participants perform a specific test, detailed in \cite{sun2019towards} to validate the transaction. After validation, the valid transactions are collected in a block of valid requests, as shown in Fig. \ref{fig3}.
\subsubsection{Quantum block proposal and validation phase} After the transaction validation phase, the quantum block (QB) of valid requests is created and broadcasted to peer nodes in the quantum blockchain network for validation. The QB is validated by using quantum-secured consensus protocols consists of proposing phase, voting phase, and decision phase, explained in \cite{sun2019towards}. When QB is validated, then it is added to the chain of QB to form a quantum-secured blockchain. After that, the ledger of each node in quantum blockchain network is updated, and transaction is securely received.

\section{Security in Optical Networks using Quantum-secured Blockchain}
In this section, we explain the distributed quantum-secured blockchain based optical network architecture for future research to enhance security of optical networks.

\subsection{Distributed Quantum-secured Blockchain Optical Network Architecture}
\begin{figure*}[ht]
    \centering
    \includegraphics[scale=.58]{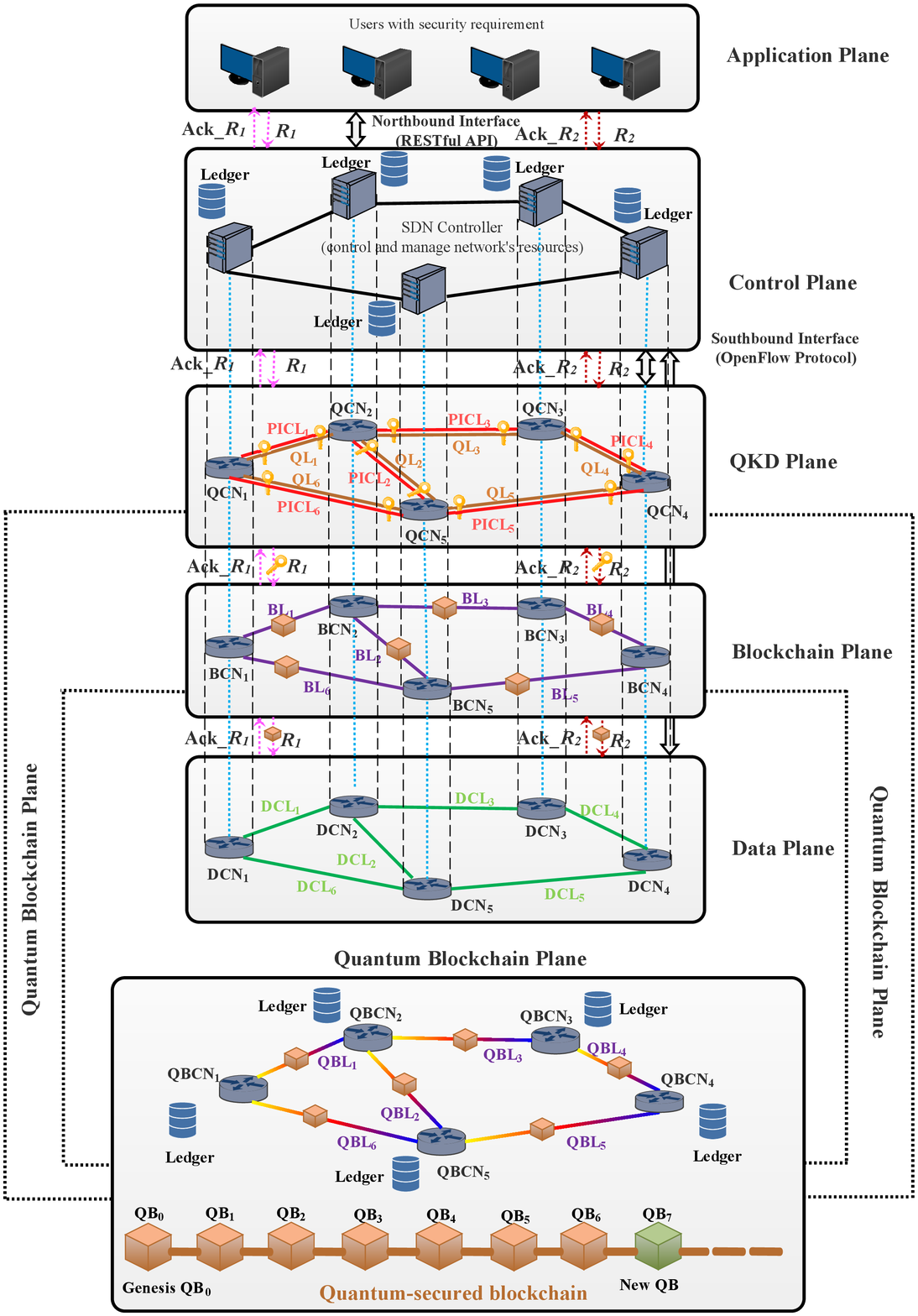}
   \caption{Distributed Quantum-secured Blockchain based Optical Network Architecture}
   \label{fig4}   
\end{figure*}

Fig. \ref{fig4} shows the distributed quantum-secured blockchain optical network architecture. The architecture consists of five planes, namely; application plane, control plane, QKD plane, blockchain plane, and data plane. The description of each plane with an example is discussed in this section and shown in Fig. \ref{fig4}.
\subsubsection{Application Plane}Application plane generates lightpath requests and sends them to the control plane.
\subsubsection{Control Plane}In this distributed network architecture, the control plane is implemented by using SDN controllers. The SDN controller efficiently controls and manages the network's resources. After generation of lightpath requests from the application plane, the control plane alert the QKD plane, blockchain plane, and data plane. Control plane allocates resources for QSCh and PICh in QKD plane, blockchain channel (BCCh) in blockchain plane i.e., for quantum blockchain channel (QBCh) in quantum blockchain plane and TDCh in the data plane.  
\subsubsection{QKD Plane} The QKD plane is implemented using the BB84 protocol for secret key generation between the end-users, as discussed in Section III. The generated secret keys through QSCh and PICh is used for blockchain security in the blockchain plane.
\subsubsection{Blockchain Plane}The blockchain plane generates different blocks of chain using secure quantum technology. The generated blockchain facilitates the process of recording and tracking requests without the need of any single centralized trusted authority. This plane helps in maintaining a ledger at each node, hence it is tamper-resistant. 
\subsubsection{Data Plane}Data plane serves the lightpath requests in a similar way as data transmission in conventional optical networks however with added security. After each secure transmission, a ledger of all nodes in the quantum-secured blockchain optical network is updated.

Consider a lightpath request \textit{R\textsubscript{1}} arriving in the network with the security requirement. Upon receiving the lightpath request from the application plane, a distributed control plane alerts the QKD plane to generate and provide requested secret keys between the nodes in the network via quantum links (QLs) and public interaction channel links (PICLs). The control plane then transmits control to the blockchain plane to securely transferring the blocks of information using QKD through blockchain links (BLs) following the process discussed in Section II (part A(2)). Combination of the QKD plane and blockchain plane of the network architecture is known as the quantum blockchain plane where the information is in the form of quantum blocks transferred through quantum blockchain links (QBLs). At the end, the data plane provides end-to-end transport of lightpath requests between the data communication nodes (DCNs) through data channel links (DCLs) in the network. After lightpath request establishment, the data plane acknowledges the control plane. The control plane after acknowledgment, updates the network's resources, ledger of each node, and status of lightpath requests generated from the application plane. In this way, a lightpath request is securely established using quantum-secured blockchain technology in the network.

\section{Research Challenges and Opportunities}
Developing new quantum-secured consensus protocols to enhance the blockchain security in optical networks is a challenging problem for future research. In addition to this, different quantum-secured signature schemes have to be designed for the signing phase in quantum-secured blockchain for authentication. Moreover, to check validity of the transaction, participants in quantum-secured blockchain have to performed a verification test. Hence, various verification schemes are needed to be proposed, which make quantum blockchain more secure and reliable for quantum-secured blockchain participants. Furthermore, one of the most critical challenges in the quantum-secured blockchain optical networks is resilience against node/link failure. Therefore, survivability needs to be addressed in such optical communication networks. Since blockchain technology is used in variety of applications such as but not limited to Internet of Things (IoT), wireless communication networks, healthcare networks, financial systems, supply chain, and voting systems. Hence, for security improvement, the quantum-secured blockchain can be deployed for such applications in the future.

\section{Conclusion}
Vulnerabilities affect optical network infrastructure and services developed for highly secure bandwidth-hungry applications such as military, finance utilities, and other government services, and cause a large amount of data and revenue loss. Hence, blockchain technology has been adopted to securely transmit the data between untrusted nodes in the optical networks. However, blockchain is vulnerable once quantum computers become easily available. Hence, quantum technology based solutions can provide opportunities to secure blockchain networks. A brief overview of blockchain and QKD is discussed with their underlying process. The article then discusses the reason behind integrating QKD into blockchain to design quantum-secured blockchain. A general distributed quantum-secured blockchain optical network architecture is presented. The architecture describes the operation of each plane to develop secure and trusted optical networks for highly secure applications against various attacks in future research. This article will raise interest towards enhancing security in optical networks and various blockchain based applications using quantum-secured blockchain.


%

%
%
%
\section*{Acknowledgment}
This work was supported by the Ministry of Education (MoE) Government of India, Ministry of Electronics and Information Technology (MeitY), and Indian Institute of Technology Indore, India.


\ifCLASSOPTIONcaptionsoff
  \newpage
\fi

\nocite{*}
\bibliography{References}
\bibliographystyle{IEEEtran}

\begin{IEEEbiographynophoto}{Purva Sharma}
is currently pursuing Ph.D. in Electrical Engineering at the Indian Institute of Technology (IIT) Indore,
India. Her research interests include WDM optical networks, elastic optical
networks, quantum key distribution, blockchain, and optical network security. \end{IEEEbiographynophoto}
\begin{IEEEbiographynophoto}{Vimal Bhatia}
is currently a Professor at Indian Institute of Technology (IIT) Indore, India. He received his Ph.D. degree from Institute for Digital Communications at The University of Edinburgh (UoE), UK in 2005 fully funded by UK Engineering and Physical Science Research Council (EPSRC) and The University of Edinburgh. During Ph.D., he also received IEE fellowship for collaborative research on OFDM with Prof. Falconer at Department of Systems and Computer Engineering at Carleton University, Ottawa, Canada. He has 13 filed patents of which 7 has been published and has published more than 260 peer-reviewed journal and conferences articles. His research focuses on optical communication networks, wireless communications, signal processing with applications to telecommunications and 5G/6G communication systems 
\end{IEEEbiographynophoto}
\begin{IEEEbiographynophoto}{Shashi Prakash}
is Professor in Department of Electronics and Instrumentation, Institute of Engineering \& Technology (IET), Devi Ahilya University,
Indore, India since 2007. He joined Devi Ahilya University in 1992. He received his M.Tech. and Ph.D. degree from Indian Institute of Technology Delhi, New Delhi, India in 1992 and 2003, respectively. He was Visiting Foreign Researcher in Department of Electrical and Electronics Engineering, Niigata University, Niigata, Japan for about a year in 2009. He has published
more than 125 journal and conference papers. His research focuses on optical communication networks, optical metrology, and laser-based instrumentation.
\end{IEEEbiographynophoto}
\end{document}